\documentstyle [12pt]{article}
\begin{document}

\hfill {WM-94-112}

\hfill {September, 1994}

\vskip 1.0in   \baselineskip 24pt {
\Large

   \bigskip
   \centerline{Solar Neutrino Oscillations in the Moon} }
 \vskip 1.0in

\centerline{Brian Mason\footnote{Current address:  Physics Department,
University of Pennsylvania, Philadelphia PA  19104.}
 and Marc Sher}  \bigskip
\centerline {\it Physics Department, College of William and Mary}
\centerline {\it Williamsburg, VA 23187}
\vskip 1.0in

{\narrower\narrower  Many different solar neutrino experiments detect
significantly fewer neutrinos than expected.  No known modification of
the standard solar model can account for all of the data.  The most
likely explanation is neutrino oscillations, in which many of the
electron neutrinos emitted by the core of the Sun convert into other
types of neutrinos while traversing the Sun.  This explanation will be
severely tested within the next few years.  We present a pedagogical
explanation of the phenomenon of neutrino oscillations, and then use the
results to address the question of whether the currently planned
neutrino detectors could detect neutrino oscillations in the moon during
solar eclipses.

}

\newpage

\def\beq{\begin{equation}}
\def\eeq{\end{equation}}

\section{Introduction}

A quarter of a century ago, Davis, Harmer and Hoffman
\cite{Davis} reported the results of the Homestake solar
neutrino experiment.
 They found that the flux of
neutrinos emitted by the Sun was a factor of 1/4 to 1/3 of
the expected flux.  Their result was confirmed several years ago by
 experimenters at Kamiokande, Japan\cite{Kamio}.  It implies that
either (a) the Standard Solar Model (SSM), used to predict the neutrino
flux, is wrong or else that (b) the neutrinos {\it are}
emitted, but changed into an undetectable type of neutrino
somewhere between the center of the Sun and the detector.

In the past few years, the experiments of
SAGE\cite{sage} and GALLEX\cite{gallex} have also reported
significant deficits in the solar neutrino flux.  Since the
neutrinos detected in these experiments come mostly from the
primary energy production mechanism in the Sun, there are much
smaller uncertainties in the Standard Solar Model prediction,
implying that the solution to the solar neutrino problem
lies in the neutrino physics (case (b) above) rather than
the SSM (case (a)).

How can neutrinos ``change" into an undetectable type of
neutrino?  As will be discussed in detail below, there
are three types of neutrinos, referred to as ``flavors'':
electron neutrinos, muon neutrinos and tau neutrinos.  Only
electron neutrinos can be emitted by the Sun, and only
those can be detected in solar neutrino experiments.
However, if neutrinos have mass, then the mass eigenstates
could be linear combinations of the three types of
neutrinos, and one can easily show that electron neutrinos
could transform into one of the other two types during travel
from the core of the Sun to the detector on Earth,  and thus
evade detection.  The preferred oscillation mechanism, called
the Mikheyev, Smirnov, Wolfenstein (MSW) mechanism\cite{ms,w},
gives an excellent description of the data for all four solar
neutrino experiments.  In this mechanism, neutrino oscillations
in matter differ from those in vacuum, and can be resonantly
enhanced when the neutrinos encounter particular densities; the
solar neutrinos then encounter these densities on their way out
of the Sun and, possibly, also on their way through the Earth
(at night).

Many particle physicists believe that the most exciting
results in particle physics during the next few years will come
from neutrino experiments.  The Solar Neutrino
Observatory in Sudbury, Canada, will begin operation next
year and will be the first experiment which is sensitive
to muon and tau neutrinos (detection of such neutrinos,
since they cannot be produced in the Sun, would be a
``smoking gun'' for neutrino oscillations).  The CHORUS
and NOMAD experiments, which are beginning to take data
now, will be sensitive to the theoretically preferred
values for muon-tau neutrino oscillations.
Super-Kamiokande will begin operation within two years,
and there are already various hints of neutrino
oscillations from atmospheric neutrino measurements.
Confirmation of neutrino oscillations would be of
enormous importance to particle physics, and would be the
first evidence of physics outside the standard model in
two decades.

Our purpose in this article is twofold.  First, the basic
physics of neutrino oscillations does not require any
more understanding than junior level quantum mechanics,
and we hope to provide an explanation of neutrino
oscillations which will enable an undergraduate to follow
the exciting imminent developments in the field.
Secondly, we wish to propose and analyze the remarkable
possibility that future neutrino experiments might be
able to detect changes in the solar neutrino flux during
a solar eclipse, due to resonant MSW oscillations in the
moon.  In Section II, we will provide some general
background, as well as a discussion of the experiments
used to measure the solar neutrino flux.  Sections III
and IV will deal, respectively, with the theory of
neutrino oscillations in vacuum and with the resonant
enhancement in matter.  In Section V, the effects of a
non-constant density medium will be analyzed.  The
calculation of the influence of the moon on the solar
neutrino flux will be discussed in Section VI.  Finally,
Section VII will contain our conclusions and a brief
discussion of the future prospects of such an experiment.

\section{Background}

The existence of the neutrino was first postulated by
Pauli in 1931 to explain the continuous energy spectrum
of electrons in beta decay.  Direct evidence for
neutrinos, however, did not materialize until 1953 when
Reines and Cowan exposed a sample of hydrogen in an
intense flux of antineutrinos from a nuclear reactor and
observed the consequent stimulated beta decay of the
protons.

Today, we know that neutrinos are electrically neutral,
possibly massless, spin 1/2 particles which interact very
weakly with ordinary matter.  A typical solar neutrino
will have a mean free path in lead of approximately a
light year, and thus neutrino detectors must look at a
large flux of neutrinos in a very large detector, in
order to have a reasonable probability of observing a
reaction.  There are three different types of neutrinos,
corresponding to the three generations of massive
leptons--the electron, the muon and the tau.
Interactions involving the electron  neutrino, for
example, only involve the electron.  Since the
temperature of the Sun is not high enough to produce
muons or taus, only electron neutrinos are emitted by the
core of the Sun.

As soon as it became clear that the primary energy
production mechanism in the Sun was nuclear fusion, it
was recognized that neutrinos would be emitted in the
process.  There are several reaction  chains which
convert hydrogen into helium in the core of the Sun.  The
proton-proton chain, which accounts for most of the
energy production, produces very low energy neutrinos.
Other chains produce different neutrino spectra; the
highest energy neutrinos arise from the conversion of
${}^7 Be$ to ${}^8B$ in the boron cycle (which only
accounts for about $10^{-4}$ of the energy production).
Predictions of the solar neutrino flux are generated from
a complicated numerical simulation of the Sun known as
the standard solar model (SSM).  The model itself is
derived from some fairly simple assumptions (such as
hydrostatic equilibrium), although the computer codes
used to generate the predictions can be quite
complicated.  The current SSM is a dynamical theory that
agrees with many known facts about the Sun, such as
spectral data, helioseismological data, and its current
mass, radius and luminosity.  The model predicts not only
the flux of solar neutrinos, but the energy spectrum as
well.

Two processes are used to observe the solar neutrino
flux: neutrino absorption and neutrino-electron
scattering. The neutrino absorption experiments all use
some form of the stimulated beta decay reaction $\nu_e\
+\ {}^AZ\rightarrow\ {}^A(Z+1)\ +\ e^-$, where $Z$ is the
charge of the nucleus and $A$ is the mass number.  Note
that only electron neutrinos can be detected in this
manner (the energy of solar neutrinos is in the MeV
range, too small to produce muons or taus).  The
solar neutrino experiment of Davis, et al., at Homestake,
South Dakota, is based on the reaction
$\nu_e+{}^{37}Cl\rightarrow\ {}^{37}Ar\ +\ e^-$.    Since the energy
of the ${}^{37}Ar$ is higher than that of the ${}^{37}Cl$, only
sufficiently high energy neutrinos can initiate this reaction.  It
turns out that the threshold for this reaction is so high that only the
high energy neutrinos from the rare boron cycle can be detected.  The
SAGE and GALLEX experiments use the conversion of gallium into
germanium.  This reaction has a much lower threshold, and thus these
experiments are sensitive to the neutrinos coming from the primary
proton-proton chain.  In all of these experiments, the final nuclear
product is radioactive.  The apparatus is allowed to stand for a long
period of time--several months for the Homestake experiment--and then
the radioactive atoms are chemically separated out from the detector and
counted in proportional counters.  Thus, they have
a time resolution on the order of months (and would
obviously be useless for looking for a change in the flux
during an hour-long solar eclipse).

One exception will be the Sudbury Neutrino Observatory
(SNO) which will observe  radiation from the produced
electrons in the reaction $\nu_e+d\rightarrow\ p+p+e^-$
and thus obtain real-time measurements of the electron
neutrino flux, direction and energy spectrum.  It is
expected to have a very high count rate (on the order of
one event per hour).

The second type of reaction used to observe solar
neutrinos is neutrino-electron scattering, or
$\nu+e\rightarrow \nu+e'$.  The Japanese experiment
at Kamiokande uses this reaction; SNO will also use this
reaction in addition to the above absorption reaction.
Typically, the detector material is water. Phototubes are used to detect
the radiation from the scattered electrons.  The rate for this reaction
is lower than for neutrino absorption; an upgrade of Kamiokande
should have several events per day.

The Homestake
experiment detects 1/4 to 1/3 the expected
flux; Kamiokande sees a similar deficit.  The SAGE and
GALLEX experiments, sensitive to a different energy
range, see approximately 60 percent of the expected rate.
There have been attempts to modify the standard solar
model to account for these discrepancies, but these
attempts have run into serious problems.  For example,
attempts to make the core temperature lower (which would
drastically cut the neutrino production rate) also lower
the luminosity too much, etc. No solar model has been
successful in explaining the solar neutrino fluxes
and still maintaining agreement with the observed mass,
radius, luminosity and helioseismology data.

If modifying the model of the Sun can't explain the solar
neutrino deficit, then one must modify the physics of
neutrinos.  The most attractive such modification
involves neutrino oscillations.  The ``electron neutrino",
$|\nu_e\rangle$, is {\it defined} to be the state which, in weak
interactions, is always associated with electrons;
similar definitions apply to the muon and tau neutrinos.
If neutrinos are massless (or have the same mass), then
there is no other way to distinguish between the three
states.  However, suppose neutrinos have mass.  Then
there will be three states
$|\nu_1\rangle,|\nu_2\rangle,|\nu_3\rangle$ which have masses
$m_1,m_2,m_3$.   How are the states $|\nu_e\rangle,|\nu_\mu\rangle,
|\nu_\tau\rangle$ related to
$|\nu_1\rangle,|\nu_2\rangle,|\nu_3\rangle$?  Each set forms a complete
description of the three neutrinos.  If one had a detector which made a
``measurement" of the masses, then the ``mass basis",
$|\nu_1\rangle,|\nu_2\rangle,|\nu_3\rangle$, would provide a
description of the possible eigenstates of the measurement; if one had
a detector which measured the weak interactions of the neutrinos (such
as a solar neutrino detector), then the ``flavor basis",
 $|\nu_e\rangle,|\nu_\mu\rangle, |\nu_\tau\rangle$, would provide a
description of the eigenstates of the measurement.  Since each basis
forms an orthonormal set, they can be related by a unitary
transformation.  For example (ignoring
mixing with the tau family), one would expect
\begin{eqnarray}\label{basis}|\nu_1\rangle&=&|
\nu_e\rangle\
 \cos\theta\
 +\ |\nu_{\mu}\rangle\
\sin\theta\nonumber \\ |\nu_2\rangle
&=&-\nu_e\rangle\ \sin\theta\ +\ |\nu_{\mu}\rangle\
\cos\theta\end{eqnarray}  where $\theta$ is an arbitrary angle which
must be experimentally determined.

 As we will see in the next
section, if an electron neutrino is produced in the core
of the Sun, then it propagates as a combination of $\nu_1$
and $\nu_2$ and, depending on the mass difference and
mixing angle, this combination will be different when the
detector is reached, leading to a combination of $\nu_e$
and $\nu_{\mu}$.  As the latter are not detected, there
will be a deficit in the neutrino flux.  Not surprisingly,
there will be a difference between propagation in vacuo
and propagation in the Sun or Earth (or moon), since there
is an effective index of refraction in a medium.  The two
free parameters in neutrino oscillations are $\Delta m^2$
and $\sin^22\theta$, where the former is the mass-squared
difference between the electron neutrino and the
(presumably) muon neutrino and the latter is the mixing
angle.  One of the major purposes of any solar neutrino
experiment is to constrain the possible values of these
parameters.  The data from  all existing solar neutrino
experiments have been combined by Hata and
Langacker\cite{hata}, who have narrowed the values down to the
region shown in Fig. 1.
  Future experiments will, within the
next couple of years, either narrow the region further, or rule
out neutrino oscillations as the solution of the solar neutrino
 problem.  The smoking gun may come from SNO, since the
neutrino-electron scattering reaction discussed above
will be also sensitive (with a different electron
distribution) to muon and tau neutrinos; their detection
 would definitely establish neutrino
oscillations as the solution, and would also establish
that these fundamental particles do have a mass.

\section{Neutrino Flavor Oscillations in
Vacuum}

The idea of neutrino flavor oscillations is
straightforward:  if a beam of pure electron neutrinos,
for example, is generated at some point then  as the beam
propagates through space it will acquire some small muon
neutrino component.  At some point, the muon neutrino
component reaches a maximum, after which the beam begins
to oscillate back into a pure electron neutrino state.
If the resonance condition (to be discussed later) is
satisfied, then at this maximum point the beam will
consist entirely of muon neutrinos.  If you had a
neutrino detector which you could move along the beam, you
would find that the flavor content of the beam varied
sinusoidally with your distance from the point at which the
neutrinos originated. Many experiments use precisely
such a procedure to look for neutrino oscillations
directly.  We will spend the rest of this section making
this argument quantitative.  Our discussion follows to
some extent that of Bernstein and Parke\cite{parke}.

We will consider the case of two neutrino flavors (the
extension to three flavors is trivial, and will not be
relevant in the following).  There are two wavefunctions,
$|\nu_1\rangle$ and $|\nu_2\rangle$ with masses $m_1$ and
$m_2$, respectively. Since the neutrinos from the Sun
have energies in the MeV range, and the experimental
bounds on the masses of the lightest two neutrinos are
much smaller than this, the neutrinos are moving at a
speed very close to the speed of light.  The time
evolution of these wavefunctions will be governed by the
relativistic analog of the Schr\"odinger equation (this
can be easily derived from the Dirac equation):
\begin{eqnarray}i{d\nu_1\over
dt}&=&E_1\nu_1=\sqrt{k^2+m_1^2}\ \nu_1\nonumber \\
 i{d\nu_2\over
dt}&=&E_2\nu_2=\sqrt{k^2+m_2^2}\ \nu_2\end{eqnarray} where we use
units of $\hbar=c=1$ and drop the ket symbol. This can be
written as
\beq i{d \over dt}\left({\nu_1(t)
\atop \nu_2(t)}\right)
=\pmatrix{\sqrt{k^2+m^2_1}&0\cr
0&\sqrt{k^2+m^2_2}\cr}\left({\nu_1(t)
\atop \nu_2(t)}\right)\eeq  This equation can be trivially
solved, yielding
\beq \left({\nu_1(t)
\atop \nu_2(t)}\right)
=\pmatrix{\exp[-i\sqrt{k^2+m^2_1}\ t]&0\cr
0&\exp[-i\sqrt{k^2+m^2_2}\ t]\cr}\left({\nu_1(0)
\atop \nu_2(0)}\right)\eeq
This propagator matrix can be used to determine the
wavefunctions at any time given the initial components
$\nu_1(0)$ and $\nu_2(0)$.

Since the neutrinos are ultrarelativistic, we can expand
the phases, using $\sqrt{k^2+m^2}\simeq k+ {m^2\over 2k}$.
Defining the intrinsically positive quantity (without
loss of generality, one can assume that $m_2>m_1$)
\beq\Delta m_o^2=m^2_2-m^2_1,\eeq the phases can then be
expressed as
\beq\sqrt{k^2+m^2_i}=[k+{m_2^2+m^2_1\over 4k}]\mp {\Delta
m^2_o\over 4k}\eeq
taking the minus sign for $i=1$ and the plus sign for
$i=2$.  Note, however, that the bracketed term is a
common phase which can be factored out of the propagator
matrix.  It then acts as an overall phase factor
to the entire wavefunction, and can be discarded since
overall phases are not measurable.  We are left with
\beq \vec{\nu}(t)=\pmatrix{\exp[i{\Delta m^2_o\over 4k}t]&0\cr
0&\exp[-i{\Delta m^2_o\over 4k}t]\cr}\left({\nu_1(0)
\atop \nu_2(0)}\right),\eeq or, equivalently,
\beq i{d\over dt}\left({\nu_1
\atop \nu_2}\right)=\pmatrix{-{\Delta m^2_o\over 4k}&0
\cr 0&{\Delta m^2_o\over 4k}\cr}\left({\nu_1
\atop \nu_2}\right).\eeq

To determine the flavor transition probabilities, we need
to express this result in terms of the flavor basis.  Using
Eq. 1 to relate $\nu_1$ and $\nu_2$  to $\nu_e$ and $\nu_\mu$,
we find that
\beq i{d\over dt}\left({\nu_e\atop \nu_\mu}\right)=
\pmatrix{-{\Delta m^2_o\over 4k}\cos 2\theta&
{\Delta m^2_o\over 4k}\sin 2\theta\cr
{\Delta m^2_o\over 4k}\sin 2\theta&
{\Delta m^2_o\over 4k}\cos 2\theta\cr}
\left({\nu_e\atop \nu_\mu}\right).\eeq
{}From this, it is clear that a beam which starts as an electron
neutrino beam will evolve into a beam with both electron
neutrino and muon neutrino components.  Solving this
equation (most easily done by looking at Eq. 7 in the flavor
basis) yields
\beq \left({\nu_e(t)\atop\nu_\mu(t)}\right)
=\pmatrix{\cos({\Delta m^2_o\over 4k}t)+i\cos 2\theta\sin
({\Delta m^2_o\over 4k}t)&i\sin 2\theta\sin({\Delta m^2_o\over 4k}
t)\cr i\sin 2\theta\sin({\Delta m^2_o\over 4k}t)&
\cos({\Delta m^2_o\over 4k}t)-i\sin 2\theta\sin
({\Delta m^2_o\over 4k}t)\cr}\left({\nu_e(0)\atop
\nu_\mu(0)}\right).\eeq
If one starts with a beam of one type of neutrino, then the
probability of conversion at a later time is then
\beq P_{\rm transition}(k,t)=|\langle\nu_\mu(t)|\nu_e(0)
\rangle|^2
=
\sin^2 2\theta\sin^2(
{\Delta m^2_o\over 4k}t)\eeq
If we now convert this to a distance $L=ct$, and express
$\Delta m^2_o$ in units of ${\rm eV}^2$, $L$ in units of
kilometers, and the energy of the neutrino in units of
MeV, the expression becomes
\beq P_{\rm transition}(k,t)=\sin^2 2\theta\sin^2(
1.27{\Delta m^2_o\over k}L).\eeq

There are several important features of this result.  First,
 it was assumed that the neutrinos were emitted at a point;
if the production region is in fact on the order of or
greater than the oscillation wavelength, then the second
$\sin$ term averages to $1/2$, leading to a probability
of transition given by ${1\over 2}\sin^2 2\theta$, which
is never bigger than $1/2$.  This could not account for
the signal reduction of greater than 50 percent observed
at Homestake\footnote{Of course, the errors in the
SSM and the experiment might accommodate a 50 percent
reduction; one can also mix all three neutrinos equally and
get a slightly bigger reduction.}.  Secondly, if
$\Delta m^2_o$ is very small, so that the oscillation length
is very long, then one can fine-tune so that one
astronomical unit is an integral number of oscillation
lengths (the integer can not be too large, since no significant
seasonal variation is seen); this requires a
 $\Delta m^2_o\simeq 10^{-10}{\rm eV}^2$.
  Even then, a very large mixing
angle is needed to get the large reduction in signal
observed by Homestake; and since the oscillation length is
energy dependent, it is difficult to also get agreement
with the SAGE and GALLEX reductions.  Thus, vacuum
oscillations are not theoretically favored and can not be made
consistent with all solar neutrino experiments.

As mentioned previously, the two parameters $\Delta m^2_o$ and
$\theta$ are free:  their values must be measured experimentally.
The details of an experiment, such as the length between the
neutrino source and the neutrino detector, the neutrino spectrum,
and systematic uncertainties, determine some region of
parameter space which may be explored. The usual procedure
is to determine the isoprobability contours for a flavor
transition and plot these in the $\Delta m^2_o$ vs. $
\sin^22\theta$
plane (or MSW plane).  For example, if a rate of five events
per hour is predicted, and a flux of one per hour is observed,
then all regions of the plane which predict anything other than
an 80 percent transition probability are excluded.  The
resulting line is then smeared out into a region by experimental
and theoretical uncertainties.  In most laboratory
experiments to date, no reduction is seen, so one plots limits
in the MSW plane.

In the next section, we will see that the interactions of
neutrinos with matter can significantly enhance flavor oscillations,
leading to possibly large transition probabilities even if
the vacuum mixing ($\sin\theta$) is small.

\section{Neutrino Flavor Oscillations in Matter}

Wolfenstein pointed out in 1978 that when a neutrino propagates
through matter,  it will interact
via the weak force with the surrounding electrons.  Both
electron and muon neutrino states are capable of scattering
off an electron (or a proton or neutron) by exchanging a
$Z^o$; this is called the weak neutral current interaction.
The electron neutrino, however, can additionally interact via
the weak charged current by exchanging a $W^+$, as shown in
Fig. 2. The net effect of these interactions is
to slow down the neutrinos' propagation---with the electron neutrino
states being slowed down more than the muon neutrino states due to the
extra mode of interaction available to them.  This additional phase
difference can enhance or repress flavor oscillations.  A beautiful
discussion of this effect can be found in a paper of Bethe\cite{bethe}.

In this section we will discuss the case of neutrino
propagation through a medium of constant density.  In the
following section we will outline the extension of this theory
to a medium of slowly changing density (the adiabatic case)
and a medium whose density changes very rapidly (the
non-adiabatic case).

First, consider the effect of the weak neutral current on the
propagator matrix in Eq. 9.  Since the weak neutral current
acts equally on the electron and muon neutrino states, the
term which appears in each diagonal element resulting from the
weak neutral current Hamiltonian is the same.  If a term
proportional to the identity matrix is added to the
Hamiltonian, then it will end up factoring out of the solution
and once again contribute an overall phase to the
wavefunction, thus having no effect on oscillations.  Thus,
the weak neutral current will not affect neutrino oscillations.

The weak charged current, on the other hand, acts {\it only}
on electron neutrino states, so there will be an added term
only in the upper left element of the Hamiltonian.  This term
is \beq {\sqrt{2}}G_FN_e\eeq  where $G_F$ is the Fermi
coupling constant which characterizes the strength of the weak
interaction and $N_e$ is the electron number density in the
medium.  Eq. 9 now looks like
\beq i{d\over dt}\left({\nu_e(t)\atop\nu_\mu(t)}\right)=
\pmatrix{ -{\Delta m^2_o \over 4k}\cos 2\theta+{\sqrt{2}}G_fN_e& {\Delta
m^2_o \over 4k}\sin 2\theta\cr {\Delta m^2_o \over 4k}\sin 2\theta&
{\Delta m^2_o \over 4k}\cos
2\theta\cr}\left({\nu_e\atop\nu_\mu}\right).\eeq
Again, adding a term proportional to the identity matrix, one
can write this as
\beq i{d\over dt}\left({\nu_e(t)\atop\nu_\mu(t)}\right)=
\pmatrix{ -{\Delta m^2_o \over 4k}\cos 2\theta+{\sqrt{2}\over
2}G_fN_e& {\Delta m^2_o \over 4k}\sin 2\theta\cr {\Delta m^2_o
\over 4k}\sin 2\theta& {\Delta m^2_o \over 4k}\cos
2\theta-{\sqrt{2}\over 2}G_fN_e
\cr}\left({\nu_e\atop\nu_\mu}\right).\eeq
This can be written in a form identical to Eq. 9 by
defining the quantities:
\begin{eqnarray}{\Delta m^2_N \over 4k}\cos 2\theta_N&=&
{\Delta m^2_o \over 4k}\cos 2\theta-{\sqrt{2}\over 2}G_FN_e
\nonumber
\\ {\Delta m^2_N \over 4k}\sin 2\theta_N&=&{\Delta m^2_o \over
4k}\sin 2\theta.\end{eqnarray}
Thus, the entire discussion of oscillations in matter is
identical to that of oscillations in vacuum with the
replacement of $\Delta m^2_o$ and $\sin 2\theta$ with
$\Delta m^2_N$ and $\sin 2\theta_N$.  Solving the above
equation explicitly gives
\begin{eqnarray} \tan 2\theta_N&=& {1\over \cot
2\theta-{2k\over \Delta m^2_o}{\sqrt{2}G_FN_e\over \sin 2\theta}
}\nonumber \\ \Delta m^2_N&=&\Delta m^2_o{\sin 2\theta\over
\sin 2\theta_N}.\end{eqnarray}
One can immediately see that if the electron density has the
value such that the denominator of the equation for $\tan
2\theta_N$ vanishes, then one can have a huge mixing angle (45
degrees) even if the vacuum mixing angle is very small.  This
is the origin of the enormous enhancement one can have for
oscillations in matter as opposed to vacuum.  Since the
electron density in the Sun varies from a large value at the
core to a small value at the surface, it is possible that that
critical electron density will be reached on the way out of
the Sun.

Since the form of the propagator matrix is identical to that of Eq. 9
(with the replacements of Eq. 16), we have the flavor transition
probability
\beq P_{\rm transition}=\sin^22\theta_N\sin^2(1.27{\Delta
m^2_N\over k}L)\eeq
The dependence of $\theta_N$ on the various parameters is
shown in Fig. 3 for the case that the vacuum mixing angle
$\theta=0.1$. Note that the angle does, as stated in the previous
paragraph, reach $45$ degrees for some value of the electron
density.  As the vacuum angle gets smaller, the width of the peak
in Fig. 3 will decrease, but the height will always be at $45$
degrees.
  Maximal mixing between flavor states occurs
when $\theta_N=\pi/4$; this resonance condition is
\beq{\Delta m^2_o\cos 2\theta\over 2kN_eG_F\sqrt{2}}=1.\eeq
It is interesting to note that far above resonance--as, for
instance, in a medium of very high density--the mixing angle
goes to zero.  Notice also that if the vacuum mixing angle is
negative, then the matter mixing angle always lies between
zero and the vacuum value, tending towards zero as the density
of the medium increases.

A detailed discussion of the transition of neutrinos in the
Sun will be given in the next section, which deals with a
non-constant density medium.  The moon has a fairly constant
density, and thus the results of this section will be
sufficient to determine the isoprobability contours for
neutrino flavor transition as they pass through the moon
during a solar eclipse.  For a lunar density of $3400\ {\rm
kg/m}^3$ (roughly that of the Earth's mantle), and for momenta
on the order of that of solar neutrinos ($O(1)$ MeV),  then
the resonance condition for small vacuum mixing will be
satisfied for $\Delta m^2_o\simeq 10^{-6} {\rm eV}^2$, which
is also, as we can see from Fig. 1, the range of masses
allowed by solar neutrino experiments.  Thus, resonant
oscillations in the moon are not implausible.  A detailed
analysis of this effect will follow in Section 6.

 There has also been much
discussion\cite{hata}
 of the possibility of matter oscillations in the
Earth (which also has a non-constant density).  Electron
neutrinos generated in the solar core can be converted to muon
neutrinos in the Sun, and can then reconvert back to electron
neutrinos in the Earth. Since only solar neutrinos detected at
night have passed through the Earth, this is called the
day/night effect, and the fact that some experiments do not yet
see a significant day/night has been used to place some
constraints in the MSW plane.  It is amusing to think that, if
the day/night effect is observed, then the Sun (observed in electron
neutrinos) could be brighter at night!!

\section{The Matter Effect in a Non-Constant Density Medium}

If the medium in which the neutrinos propagate  does not
have a constant density, then we must be much more careful.
This case is especially relevant to the solar neutrino problem
because the Sun's density varies strongly between the core and
the photosphere.  It is in fact roughly exponential, with a
maximum density of $150,000\ {\rm kg/m}^3$ at the center.

The case of a slowly changing density profile is called
adiabatic since the matter mass eigenstates will vary smoothly
(i.e., without mixing between these states) into the vacuum
mass eigenstates as the density goes to zero.  If the density
profile changes rapidly, then the mass eigenstates may be
mixed at this point.  The technical condition for adiabaticity
is somewhat subtle; see the book of Bahcall\cite{bahcall} for
details.  We will first assume that the {\it matter mixing angle}
changes slowly, which one can show, from the adiabaticity
condition, is satisfied except at a resonance crossing (see
Fig. 3).

The adiabatic evolution of a neutrino state is determined by a
simple generalization of the results of the previous
section:\beq i{d\over dt}\left({\nu_e\atop\nu_\mu}\right)=
\pmatrix{-{\Delta m^2_o \over 4k}\cos 2\theta+{\sqrt{2}\over 2}
G_FN_e(t)&{\Delta m^2_o \over 4k}\sin 2\theta\cr
{\Delta m^2_o \over 4k}\sin 2\theta&
{\Delta m^2_o \over 4k}\cos 2\theta-{\sqrt{2}\over 2}G_FN_e(t)
\cr}\left({\nu_e\atop\nu_\mu}\right).\eeq
This differs only in the time dependence of the electron
number density.  One can then define the time dependent mass
and mixing angle as
\begin{eqnarray} {\Delta m^2_N(t) \over 4k}\cos
2\theta_N(t)&=& {\Delta m^2_o \over 4k}\cos 2\theta-
{\sqrt{2}\over 2}G_fN_e(t)\nonumber \\
{\Delta m^2_N(t) \over 4k}\sin 2\theta_N(t)&=&{\Delta m^2_o
\over 4k}\sin 2\theta\end{eqnarray}
With these definitions, the expressions for $\nu_e(t)$ and
$\nu_\mu(t)$ in terms of their values at $t=0$ become
identical to Eq. [16] with the time-dependence of $\theta_N$
and $\Delta m^2_N$ included.  This time-dependence, of course,
can be trivially converted into a distance, since the
neutrinos are travelling very close to the speed of light.

The non-adiabatic case is too complicated to deal with in
detail here.  We will summarize the useful results.  The
survival probability for a neutrino which propagates through a
density resonance (whose existence is established by
satisfying the resonance condition of Eq. [19] somewhere in
the Sun) is \beq P_s=\sin^2\theta + P_x\cos 2\theta\eeq
where $P_x$ is the Landau-Zener level crossing probability.
This factor gives the probability of a transition between the
adiabatic states during resonance crossing.  It is given by
\beq P_x= {\rm Exp}[-{\pi\over 2}{\sin^2 2\theta\over \cos
2\theta}{\Delta m^2_o\over 2k}{1\over {1\over N_e}{dN_e\over
dt}|_{\rm res.}}]\eeq  The logarithmic derivative in the
denominator is evaluated at the time of the resonance crossing.

The probability of detecting a neutrino in its original flavor
state after two resonance crossings is obtained by replacing
$P_x$ with $ P_{1x}(1-P_{2x})+P_{2x}(1-P_{1x})$.  The
generalization to many crossings is then simple.

Although straightforward, the detailed description of the
evolution in the solar neutrino flux as the neutrinos leave
the Sun must be done numerically, primarily because the
neutrino spectrum emitted by the core in the first place
(determined from the SSM) is fairly complicated, and all of
the expressions are energy-dependent.  Figure 4 shows the
transition probability for neutrinos generated in the core of
the Sun.
  The horizontal line at about $2\times 10^{-4}\ {\rm
eV}^2$ is due to the resonance crossing condition, Eq. [19].
Above this line, there is no resonance crossing in the Sun for
the relevant range of neutrino energies, and the vacuum
oscillations are the dominant effect.  Below this line we must
use the Landau-Zener expression above; the slope of the
diagonal lines follow from the form of $P_x$.  Of course, no
experiment is sensitive to the entire range of solar
neutrinos; so each maps out a region in the plane.  Combining
all of the experiments gives the allowed region in Fig.1.

 One
must be cautioned about taking Fig. 1 too seriously; if one of
the experiments is wrong, then the allowed regions can expand
considerably, and thus when discussing the effects of the moon
on solar neutrinos, we will consider the general region
covered in Fig. 4.  We now apply the MSW model to the
possibility of seeing a change in the solar neutrino flux
during a solar eclipse.

\section{Neutrino Oscillations in the Moon}

{}From the vantage point of a typical solar neutrino detector, the center
of the Sun is covered for a few hours every decade during solar
eclipses.  Although total eclipses are rare, partial eclipses occur
much more frequently, and one must remember that the Earth is
transparent to neutrino detectors.  Could neutrino oscillations in the
moon be detected?  Since there is not likely to be any systematic effect
which ``turns on" only during eclipses, one can integrate over a large
number of eclipses.  We will now analyze the possibility of observing a
change in the solar neutrino rate during a solar eclipse.

Specifically, one calculates the percent change in the eclipsed signal
relative to the uneclipsed signal for many points in the MSW plane, and
plots the resulting contours in this plane.  By ``signal", we mean the
total number of expected solar neutrino events during the time of
observation (on the order of an hour per eclipse).  There are two
possible effects that the moon can have on the signal.  Neutrinos which
survive their passage through the Sun as electron neutrinos can convert
into muon or tau neutrinos in the moon, and neutrinos which transform
on their way out of the Sun can reconvert in the moon.  The latter
effect will turn out to be more significant, thus the effect of the
moon is to {\it increase} the electron neutrino flux.

To find the expected uneclipsed signal for given values of $\Delta
m^2_o$ and $\sin^22\theta$, we first calculate the event rate of the
detector and then integrate this over the duration of the eclipse.  The
rate is obtained from the integrals
\beq P=\Sigma_{i={e,\mu,\tau}}\int\ \sigma_i(k)\phi_i(k)\ dk\eeq
where $\phi_i(k)$ is the spectral flux of neutrino type $i$ arriving at
the detector and $\sigma_i$ is the cross-section for a neutrino
interacting with a detector atom.  For the reaction we are considering,
$\nu_e+d\rightarrow p+p+e^-$ (the primary reaction for SNO),
$\sigma_\mu$ and $\sigma_\tau$ both vanish; for the reaction
$\nu_i+e^-\rightarrow\nu_i+e^-$, the values for $\sigma_\mu$ and
$\sigma_\tau$ are about a factor of seven less than $\sigma_e$.  The
cross sections for the nuclear reaction are obtained from extrapolating
know nuclear cross-sections to low energies.  We used the nuclear
cross-sectons of Bahcall\cite{bahcall}.

The spectral neutrino fluxes at the detector are obtained by
propagating the intial $\nu_e$ signal through the sun and the vacuum
using the techniques described in the previous three sections.  We have
\begin{eqnarray}\phi_i(k)&=&\phi_e^o|\langle\nu_i(t)|\nu_e(0)\rangle|^2 \\
&=&\phi_e^oP_{e\rightarrow i}(k)\end{eqnarray}
where $P_{e\rightarrow i}$ is the probability of an electron neutrino
generated at the solar core being detected as an $i$-neutrino after
propagating to the earth and $\phi_e^o$ is the initial neutrino flux
from the standard solar model.  For the deuteron break-up reaction at
SNO,
$i=e$ and thus this becomes, in more familiar terminology,
\beq \phi_e(k)=\phi_e^o(k)P_s(k)\eeq
If there is no resonance crossing for a particular energy and
combination of $\Delta m^2_o$ and $\theta$ on the way out of the sun,
then $P_s$ is obtained from Eq. 12, appropriately averaged over the
production position.  If there is a resonance density crossing in the
Sun, then $P_s$ comes from Eq. 22.

The eclipsed signal is obtained in a similar manner, i.e. by
integrating the rate over one eclipse, with the only difference being
that the solar neutrino flux at the detector is modified by the
intervention of the moon.  This flux will also have some
time-dependence due to the time dependence of the neutrino path length
through the moon.  The extreme non-adiabatic transition to the lunar
medium (which may be thought of as a measurement) separates the states
and allows one to ignore interference terms.  Hence we use the
classical probability result
\beq \phi_e(k,t)=\phi_e^o(k)[P_{s,sun}(k)P_{s,moon}(k,t)
+P_{t,sun}(k)P_{t,moon}(k,t)]\eeq
for the electron neutrino flux at the detector.  $P_{(s,t)}$ specify
the probability of survival (transition) for a flavor eigenstate
propagating through the given medium.  The probabilities for the moon
are from Eq. 18 with the electron density appropriate to the moon
(about $10^{30}{\rm m}^{-3}$).  In this case
\beq L(t)=2(R^2-b^2-(vt)^2)^{1/2}\eeq
Here, $R$ is the radius of the moon, $v$ is the velocity of the moon
relative to the sun as seen from the earth (about $1$ km/sec) and $b$
is the impact parameter of the eclipse (zero for a total eclipse).  We
have set $t=0$ at the eclipse maximum. Since the density of the moon
is (to an accuracy of ten percent or so) constant, Eq. 18 will be valid
for all energies and regions of the parameter space; i.e. there is no
``resonance crossing" as for the Sun.  There is, however, a resonance
effect due to the length of oscillations in the lunar medium becoming
comparable to the lunar diameter.  This enhances flavor oscillations
strongly in the vicinity of $\Delta m^2_o=5\times 10^{-6}{\rm eV}^2$,
as shown in Fig. 5, in which we plot the flavor transition probability
for the moon in the same manner as done for the Sun in Fig. 4.  Note that
this is a region of parameter space which is interesting to us (see
Fig 1).

\section{Results and Conclusion}

We calculated the total percentage change in the electron neutrino
signal which would be observed through $\nu_e + d\rightarrow p+p+e^-$
at the Sudbury Neutrino Observatory, for 900 points in the MSW plane.
Fig. 6 is the result. The
figure may be intuitively understood by imagining the ``multiplication"
of Figs. 4 and 5 (remembering that these figures are for a
specific neutrino energy, whereas for Fig. 6 we have integrated
over the spectrum).
 For example, note the large peak at
$\sin^22\theta\simeq 0.4,
\ \Delta m^2_o\simeq 7\times 10^{-6}{\rm eV}^2$ for which the flux
is increased by a factor of $7$!  As one can see from Fig. 4, however,
for these parameters over $90$ percent of the electron neutrinos
emitted by the core of the Sun are converted within the Sun; the
factor of $7$ simply means that some number of these are converted
back.  This region is already excluded, as can be seen in Fig. 1.
Comparing with Fig. 1, we can see that for the small angle
MSW solution, there is no measureable effect from the moon.
However, for the large angle solution, there is an enhancement in
the signal which ranges from 50 to 100 percent.

The calculation is presented with impact parameter $b=0$.  It was
repeated with impact parameters $0.25R,\ 0.5R$ and $0.75R$, and the
dependence of Fig. 6 was found to be weak.  This is not surprising due
to the square-root dependence of the maximum neutrino path length in
the moon on the impact parameter.

With the current generation of detectors, statistically significant
data is an impossibility for this measurement.  However, the coming
generation detectors SNO, Super-Kamiokande, Icarus and Borexino will
all have rates on the order of one event per hour\cite{revol}.  Adding
the data from multiple eclipses will provide a larger sample.  Over the
next 15 years, there will be roughly five hours during which the center
of the Sun is covered at all of these sites\cite{kay}.  At Borexino, for
example, one would expect about 10 events during these time periods (and
many more in the next round of detectors).  From Fig. 6, one can see that
there is a region in the MSW plane in which a statistically significant
detection of the moon's solar neutrino shadow is possible.

In this article, we have shown how neutrino oscillations can explain
the observed deficit in solar neutrinos.  Even with a very small mixing
angle, oscillations can be resonantly enhanced as the neutrinos travel
from the core of the Sun to the surface.  In addition to a pedagogic
review, we have examined the possibility that the solar neutrinos could
undergo further transformation if they pass through the moon during a
solar eclipse.  There is a region in the MSW plane in which a change in
the neutrino flux during an eclipse could be observed in the next
generation of solar neutrino detectors.

This work constituted the undergraduate senior thesis of Brian Mason.
We thank Peter Kay for determining the number of hours of eclipse at
each neutrino detector site.  This work was supported by the National
Science Foundation.

\def\prd#1#2#3{{\it Phys. ~Rev. ~}{\bf D#1}, #3 (19#2)}
\def\plb#1#2#3{{\it Phys. ~Lett. ~}{\bf B#1}, #3 (19#2)}
\def\npb#1#2#3{{\it Nucl. ~Phys. ~}{\bf B#1}, #3 (19#2)}
\def\prl#1#2#3{{\it Phys. ~Rev. ~Lett. ~}{\bf #1}, #3 (19#2)}

\bibliographystyle{unsrt}

\begin{thebibliography}{99}
\bibitem{kay}
This was obtained by using ``Dance of the Planets" to find the times of
all eclipses (partial and total) over the next fifteen years, and
watching the Sun (through a transparent earth, if necessary) from each
of the solar neutrino detector sites.  The center of the Sun is covered
for roughly five hours from each site during that period.
\end{thebibliography}

\newpage

\section*{Figure Captions}

\begin{description}
\item[] Fig. 1 -
The regions of the $\Delta m^2-\sin^22\theta$
plane allowed by combining all of the solar neutrino experiments to date,
assuming that solar neutrino oscillations are the reason for the
observed deficit.  This figure is Fig. 3 from Ref. 8; we thank
Paul Langacker for providing us with a copy of the figure.
\item[] Fig. 2-
Interactions of neutrinos with electrons.
All neutrinos interact with electrons via $Z^o$-exchange, but only
electron neutrinos can interact via $W^+$ exchange.
\item[] Fig. 3-
The matter mixing angle, assuming a vacuum mixing
angle of
$0.1$, as a function of $\beta$, where $\beta=
{2\sqrt{2}G_FN_e\over \sin 2\theta}{k\over \Delta m^2_o}$.  We have
plotted $|\theta_N|$ since only $\sin^22\theta_N$ is relevant;
$\theta_N$ is negative to the right of the peak.
\item[] Fig. 4-
Transition probablity for electron neutrinos
travelling from the core of the Sun to the surface.  Contour intervals
are $10$ percent.  We have taken the neutrino energy to be $7$ MeV for
this plot; different energies will yield similar curves.
\item[] Fig. 5-
Transition probability for electron neutrinos
travelling through the moon.  As in Fig. 4, the neutrino energy is taken
to be $7$ MeV.  Contour intervals are $10$ percent.
\item[] Fig. 6-
Total change in the neutrino flux observed at SNO
during a solar eclipse. We have plotted the fractional change (so that
$100$ percent corresponds to a factor of two increase in the flux).
The contours, starting with the outer contour and working inward,
are 5, 50, 100, 200, 400 and 600 percent, respectively.
\end{description}

\end{document}